\documentclass[%
 reprint,%
 amssymb, amsmath,%
 aip,cha,%
]{revtex4-1}

\usepackage{docs}%
\usepackage{bm}%
\usepackage[colorlinks=true,linkcolor=blue]{hyperref}%
\expandafter\ifx\csname package@font\endcsname\relax\else
 \expandafter\expandafter
 \expandafter\usepackage
 \expandafter\expandafter
 \expandafter{\csname package@font\endcsname}%
\fi
\begin{document}

\title{Vortex Solution of a Twisted Baby Skyrme Equation}%

\author{Malcolm Anderson$^1$,~Miftachul Hadi$^{1,2}$,~Andri Husein$^3$}%
\email{itpm.id@gmail.com (Miftachul Hadi)}
\affiliation{$^1$Department of Mathematics, Universiti Brunei Darussalam, Negara Brunei Darussalam\\
             $^2$Physics Research Centre, Indonesian Insitute of Sciences, Puspiptek, Serpong, Indonesia\\
		     $^3$Department of Physics, University of Sebelas Maret, Surakarta, Indonesia}%

\begin{abstract}
We examine non-linear sigma (plus Skyrme term) model in flat space, in particular with a twist, which comprises a twisted baby Skyrmion vortex solution with an added dependence on a twist term $mkz$, where $z$ is the vertical coordinate. We find that the solutions should have asymptotic form. We also find that mass per unit length of string, $\mu$, increases roughly linearly with $\zeta$, where $\zeta=4\lambda^2K_s(mk)^2$, $\lambda$ is scaling constant, $K_s$ is Skyrme term, $m$ is integer and $k$ is wave number.
\end{abstract}

\maketitle

\section{Introduction}
In order to understand the charge radius of nucleon which have size roughly 1 Fermi, Tony Hilton Royle Skyrme in 1962 proposed idea that strongly interacting particles (hadrons) were locally concentrated static solution of extended non-linear sigma (chiral) model. In (3+1)-dimensions of space-time, it will be observed Skyrme model which is trying to explain hadrons as solitons from non-linear sigma (chiral) model field theory in internal symmetrical group of $SU(2)$. Skyrme's idea is unifying bosons and fermions in a common framework which provide a fundamental fields model consisted of just the pion. The nucleon was obtained, as a certain classical configuration of the pion fields \cite{hans2005, mif1, sky1, sky2, val}. 

Skyrmion is a topological soliton object which is solution to the classical field equation with localized energy density \cite{mif1}. It is a classical static field configuration of minimal energy in a non-linear scalar field theory. The scalar field is the pion field and the Skyrmion represents a baryon. The Skyrmion has a topological charge which prevents it being continuously deformed to the vacuum field configuration. This topological charge is identified with the conserved baryon number which prevents a proton from decaying into pions. Mathematically, the Skyrmion is a topologically non-trivial map from physical 3-dimensional space $S$ to a target manifold $\Sigma$ \cite{manton1}. A field configuration at a given time is a map from space to the group manifold of $SU(2)$. The degree of this map, which is an integer, is identified with the physical baryon number. The Skyrmion is the map of degree 1 with minimal energy \cite{manton2}.

A nonlinear sigma model is an $N$-component scalar field theory in which the fields are functions defining a mapping from the space-time to a target manifold \cite{Zakrzewski}. 
By a nonlinear sigma model, we mean a field theory with the following properties \cite{hans02}:
\begin{itemize}
\item[(1)] The fields, $\phi(x)$, of the model are subject to nonlinear constraints at all points $x\in\mathcal{M}_0$, where $\mathcal{M}_0$ is the source (base) manifold, i.e. a spatial submanifold of the (2+1) or (3+1)-dimensional space-time manifold.
\item[(2)] The constraints and the Lagrangian density are invariant under the action of a global (space-independent) symmetry group, $G$, on $\phi(x)$.
\end{itemize}

The Lagrangian density of a free (without potential) nonlinear sigma model on a Minkowski background space-time is defined to be \cite{chen}
\begin{equation}\label{1}
\mathcal{L}=\frac{1}{2\lambda^2}~\gamma_{AB}(\phi)~\eta^{\mu\nu}~\partial_\mu\phi^A~\partial_\nu\phi^B
\end{equation}
where $\gamma_{AB}(\phi)$ is the field metric, $\eta^{\mu\nu}=\text{diag}(1,-1,-1,-1)$ is the Minkowski tensor, $\lambda$ is a scaling constant with dimensions of (length/energy)$^{1/2}$ and $\phi={\phi^A}$ is the collection of fields. Greek indices run from 0 to $d-1$, where $d$ is the dimension of the space-time, and upper-case Latin indices run from 1 to $N$.  

The simplest example of a nonlinear sigma model is the $O(N)$ model, which consists of $N$ real scalar fields, $\phi^A$, $\phi^B$, with the Lagrangian density \cite{hans02}
\begin{equation}\label{2}
\mathcal{L}=\frac{1}{2\lambda^2}~\delta_{AB}~\eta^{\mu\nu}~\frac{\partial\phi^A}{\partial x^\mu}~\frac{\partial\phi^B}{\partial x^\nu}
\end{equation}
where the scalar fields, $\phi^A$, $\phi^B$, satisfy the constraint
\begin{equation}\label{3}
\delta_{AB}~\phi^A\phi^B=1
\end{equation}
and $\delta_{AB}$ is the Kronecker delta. The Lagrangian density (\ref{2}) is obviously invariant under the global (space independent) orthogonal transformations $O(N)$, i.e. the group of $N$-dimensional rotations \cite{hans02}
\begin{equation}\label{4}
\phi^A\rightarrow\phi'^A=O^A_B~\phi^B.
\end{equation}

One of the most interesting examples of a $O(N)$ nonlinear sigma model, due to its topological properties, is the $O(3)$ nonlinear sigma model in (1+1)-dimensions, with the Lagrangian density 
\begin{equation}\label{5}
\mathcal{L}=\frac{1}{2\lambda^2}~\eta^{\mu\nu}~\partial_\mu\phi~.~\partial_\nu\phi 
\end{equation}
where $\mu$ and $\nu$ range over $\{0,1\}$, and $\phi=(\phi^1,\phi^2,\phi^3)$, subject to the constraint $\phi\cdot\phi=1$, where the dot (.) denotes the standard inner product on real coordinate space of three dimensions, $R^3$. For a $O(3)$ nonlinear sigma model in any number $d$ of space-time dimensions, the target manifold is the unit sphere $S^2$ in $R^3$, and $\mu$ and $\nu$ in the Lagrangian density (\ref{5}) run from 0 to $d-1$.

A simple representation of $\phi$ (in the general time-dependent case) is
\begin{equation}\label{6}
\phi=
\begin{pmatrix}
\sin f(t,{\bf r})~\sin g(t,{\bf r}) \\
\sin f(t,{\bf r})~\cos g(t,{\bf r}) \\
\cos f(t,{\bf r})
\end{pmatrix}
\end{equation}
where $f$ and $g$ are scalar functions on the background space-time, with Minkowski coordinates $x^\mu=(t,{\bf r})$. In what follows, the space-time dimension, $d$, is taken to be 4, and so $\bf r$ is a 3-vector.

If we substitute (\ref{6}) into the Lagrangian density (\ref{5}), then it becomes
\begin{equation}\label{7}
\mathcal{L}=\frac{1}{2\lambda^2}[\eta^{\mu\nu}~\partial_\mu f~\partial_\nu f+(\sin^2f)~\eta^{\mu\nu}~\partial_\mu g~\partial_\nu g]
\end{equation}
The Euler-Lagrange equations associated with $\mathcal{L}$ in (\ref{7}) are 
\begin{eqnarray}\label{8}
\eta^{\mu\nu}~\partial_\mu\partial_\nu f-(\sin f~\cos f)~\eta^{\mu\nu}~\partial_\mu g~\partial_\nu g=0
\end{eqnarray}
and
\begin{eqnarray}\label{9}
\eta^{\mu\nu}~\partial_\mu\partial_\nu g+2(\cot f)~\eta^{\mu\nu}~\partial_\mu f~\partial_\nu g=0.
\end{eqnarray}

\section{Soliton Solution}
Two solutions to the $O(3)$ field equations (\ref{8}) and (\ref{9}) are 
\begin{itemize}
\item[(i)] a monopole solution, which has form
\begin{equation}\label{10}
\phi=\hat{\textbf{r}}=
\begin{pmatrix}
x/\rho\\
y/\rho\\
z/\rho\\
\end{pmatrix}
\end{equation}
where $\rho=(x^2+y^2+z^2)^{1/2}$ is the spherical radius; and
\item[(ii)] a vortex solution, which is found by imposing the 2-dimensional ''hedgehog'' ansatz
\begin{equation}\label{11}
\phi=
\begin{pmatrix}
\sin f(r)~\sin (n\theta-\chi)\\
\sin f(r)~\cos (n\theta-\chi)\\
\cos f(r)
\end{pmatrix}
\end{equation}
where $r=(x^2+y^2)^{1/2}$, $\theta=\arctan (x/y)$, $n$ is a positive integer, and $\chi$ is a constant phase factor. In this thesis, we only consider the vortex solution.
\end{itemize}

A vortex is a stable time-independent solution to a set of classical field equations that has finite energy in two spatial dimensions; it is a two-dimensional soliton. In three spatial dimensions, a vortex becomes a string, a classical solution with finite energy per unit length \cite{preskill}. Solutions with finite energy, satisfying the appropriate boundary conditions, are candidate soliton solutions \cite{manton}. 

The boundary conditions that are normally imposed on the vortex solution (\ref{11}) are $f(0)=\pi$ and $\lim_{r\to\infty}f(r)=0$, so that the vortex ''unwinds'' from $\phi=-\hat{\textbf{z}}$ to $\phi=\hat{\textbf{z}}$ as $r$ increases from 0 to $\infty$. The function $f$ in this case satisfies the field equation 
\begin{equation}\label{12}
r~\frac{d^2f}{dr^2}+\frac{df}{dr}-\frac{n^2}{r}~\sin f~\cos f=0
\end{equation}
There is in fact a family of solutions to this equation (\ref{12}) satisfying the standard boundary conditions 
\begin{equation}\label{13}
\sin f=\frac{2K^{1/2}r^n}{Kr^{2n}+1}
\end{equation}
or equivalently
\begin{equation}\label{14}
\cos f=\frac{Kr^{2n}-1}{Kr^{2n}+1}
\end{equation}
where $K$ is positive constant.

The energy density, $\sigma$, of a static (time-independent) field with Lagrangian density, $\mathcal{L}$, (\ref{7}) is
\begin{eqnarray}\label{15}
\sigma 
&=& -\mathcal{L} \nonumber\\
&=& \frac{1}{2\lambda^2}\left[\eta^{\mu\nu}~\partial_\mu f~\partial_\nu f+(\sin^2 f)~\eta^{\mu\nu}~\partial_\mu g~\partial_\nu g\right]
\end{eqnarray}
The energy density of the vortex solution is
\begin{eqnarray}\label{17}
\sigma =\frac{4Kn^2}{\lambda^2}\frac{r^{2n-2}}{(Kr^{2n}+1)^2}
\end{eqnarray}
The total energy
\begin{equation}\label{18}
E=\int\int\int \sigma~ dx~dy~dz,
\end{equation}
of the vortex solution is infinite. But, the energy per unit length of the vortex solution
\begin{eqnarray}\label{19}
\mu
&=& \int\int \sigma~dx~dy=2\pi\int_0^\infty\frac{4Kn^2}{\lambda^2}\frac{r^{2n-2}}{(Kr^{2n}+1)^2}~r~dr  \nonumber\\
&=& \frac{4\pi n}{\lambda^2}
\end{eqnarray}
is finite, and does not depend on the value of $K$. (We use the same symbol for the energy per unit length and the mass per unit length, due to the equivalence of energy and mass embodied in the relation $E=mc^2$. Here, we choose units in which $c=1$). 

This last fact means that the vortex solutions in the nonlinear sigma models have no preferred scale. A small value of $K$ corresponds to a more extended vortex solution, and a larger value of $K$ corresponds to a more compact vortex solution, as can be seen by plotting $f$ (or $-\mathcal{L}$) for different values of $K$ and a fixed value of $n$. This means that the vortex solutions are what is called neutrally stable to changes in scale. As $K$ changes, the scale of the vortex changes, but the mass per unit length, $\mu$, does not. Note that because of equation (\ref{19}), there is a preferred winding number, $n=1$, corresponding to the smallest possible positive value of $\mu$.

Furthermore, it can be shown that the topological charge, $T$, of the vortex defined by
\begin{eqnarray}\label{20}
T\equiv \frac{1}{4\pi}~\varepsilon_{ABC}\int\int \phi^A~\partial_x \phi^B ~\partial_y \phi^C~dx~dy
\end{eqnarray}
where $\varepsilon_{ABC}$ is the Levi-Civita symbol, is conserved, in the sense that $\partial_t T=0$ no matter what coordinate dependence is assumed for $f$ and $g$ in (\ref{11}). 

So, the topological charge is a constant, even when the vortex solutions are perturbed. Also, it is simply shown that for the vortex solutions 
\begin{eqnarray}\label{21}
T
&=&-\frac{1}{\pi}n~[f(\infty)-f(0)]= -\frac{1}{\pi}n~(0-\pi) \nonumber\\
&=& -\frac{1}{\pi}n~(-\pi)= n
\end{eqnarray}
and so, the winding number is just the topological charge. 

Because there is no natural size for the vortex solutions, we can attempt to stabilize them by adding a Skyrme term to the Lagrangian density.  For compact twisting solutions such as the twisted baby Skyrmion string \cite{nitta1}, in addition to the topological charge, $n$, there is a second conserved quantity called the Hopf charge \cite{nitta1}, \cite{mif55}. 

\section{Skyrmion Vortex without a Twist} 
The original sigma model Lagrangian density (with the unit sphere as target manifold) is
\begin{eqnarray}\label{22}
\mathcal{L}_1=\frac{1}{2\lambda^2}~\eta^{\mu\nu}~\partial_\mu\phi~.~\partial_\nu\phi
\end{eqnarray}
If a Skyrme term is added to (\ref{22}), the result is a modified Lagrangian density
\begin{eqnarray}\label{23}
\mathcal{L}_2
&=&\frac{1}{2\lambda^2}~\eta^{\mu\nu}~\partial_\mu\phi~.~\partial_\nu\phi \nonumber\\
&&-~K_s~\eta^{\kappa\lambda}~\eta^{\mu\nu}(\partial_\kappa\phi\times\partial_\mu\phi)~.~(\partial_\lambda\phi\times\partial_\nu\phi) 
\end{eqnarray}
where the Skyrme term is the second term on the right hand side of (\ref{23}). Here, $K_s$ is a positive coupling constant.

With the choice of field representation (\ref{6}), equation (\ref{23}) becomes  
\begin{eqnarray}\label{24}
\mathcal{L}_2
&=&\frac{1}{2\lambda^2}\left(\eta^{\mu\nu}~\partial_\mu f~\partial_\nu f+\sin^2f~\eta^{\mu\nu}~\partial_\mu g~\partial_\nu g\right) \nonumber\\
&&-~K_s\left[2\sin^2f\left(\eta^{\mu\nu}~\partial_\mu f~\partial_\nu f\right)\left(\eta^{\kappa\lambda}~\partial_\kappa g~\partial_\lambda g\right) \right.\nonumber\\
&&\left.-~2\sin^2f\left(\eta^{\mu\nu}~\partial_\mu f~\partial_\nu g\right)^2\right]
\end{eqnarray}
If the vortex configuration (\ref{11}) for $\phi$ is assumed, the Lagrangian density (\ref{7}) becomes 
\begin{eqnarray}\label{25}
\mathcal{L}
&=& -\frac{1}{2\lambda^2}\left[\left(\frac{df}{dr}\right)^2 +\frac{n^2}{r^2}\sin^2f\right] \nonumber\\
&&-~2K_s\frac{n^2}{r^2}\sin^2f\left(\frac{df}{dr}\right)^2
\end{eqnarray}

The Euler-Lagrange equations generated by $\mathcal{L}_2$ (\ref{24}), namely
\begin{eqnarray}\label{26}
\partial_\alpha\left[\frac{\partial\mathcal{L}_2}{\partial(\partial_\alpha f)}\right]  -\frac{\partial\mathcal{L}_2}{\partial f}=0
\end{eqnarray}
and
\begin{eqnarray}\label{27}
\partial_\alpha\left[\frac{\partial\mathcal{L}_2}{\partial(\partial_\alpha g)}\right]  -\frac{\partial\mathcal{L}_2}{\partial g}=0
\end{eqnarray}
Reduce to a single second-order equation for $f$ 
\begin{eqnarray}\label{28}
0
&=& \frac{1}{\lambda^2}\left(\frac{d^2f}{dr^2}+\frac{1}{r}\frac{df}{dr}-\frac{n^2}{r^2}\sin f\cos f\right) \nonumber\\
&&+~4K_s~\frac{n^2}{r^2}~\sin^2f\left(\frac{d^2f}{dr^2}-\frac{1}{r}\frac{df}{dr}\right) \nonumber\\
&&+~4K_s~\frac{n^2}{r^2}~\sin f~\cos f\left(\frac{df}{dr}\right)^2
\end{eqnarray}
with the boundary conditions $f(0)=\pi$ and $\lim_{r\rightarrow\infty}f(r)=0$ as before.

If a suitable vortex solution $f(r)$ of this equation exists, it should have a series expansion for $r<<1$ of the form
\begin{eqnarray}\label{29}
f = \pi +ar+br^3+...~~\text{if}~n=1
\end{eqnarray}
or
\begin{eqnarray}\label{30}
f = \pi +ar^n+br^{3n-2}+...~~\text{if}~n\geq 2
\end{eqnarray}
where $a<0$ and $b$ are constants, and for $r>>1$ the asymptotic form
\begin{eqnarray}\label{31}
f = Ar^{-n}-\frac{1}{12}A^3r^{-3n} + ...
\end{eqnarray}
for some constant $A>0$. 

However, it turns out that it is not possible to match these small-distance and large-distance expansions if $K_s\neq 0$: meaning that any solution $f$ of (\ref{28}) either diverges at $r=0$ or as $r\rightarrow\infty$. This result follows from the following simple scaling argument.

Suppose that $f(r)$ is a solution of equation (\ref{28}). Let $q$ be any positive constant and define $f_q(r)\equiv f(qr)$. Substituting $f_q$ in place of $f$ in equation (\ref{28}) gives a value of $\mu$ which depends in general on the value of $q$
\begin{eqnarray}\label{32}
\mu_q
&=&\int\int\left\{\frac{1}{2\lambda^2}\left[\left(\frac{df_q}{dr}\right)^2+\frac{n^2}{r^2}\sin^2f_q\right] \right.\nonumber\\
&&\left.-2K_s\frac{n^2}{r^2}\sin^2f_q\left(\frac{df_q}{dr}\right)^2\right\}r~dr~d\theta
\end{eqnarray}
where
\begin{eqnarray}\label{33}
\frac{df_q}{dr} =qf'(qr)
\end{eqnarray}

So, if $r$ is replaced as the variable of integration by $\overline{r}=qr$, we have
\begin{eqnarray}\label{34}
\mu_q
&=&\int\int\left\{\frac{1}{2\lambda^2}\left[\left(\frac{df(\overline{r})}{d\overline{r}}\right)^2+\frac{n^2}{\overline{r}^2}\sin^2f(\overline{r})\right] \right.\nonumber\\
&&\left.+~2q^2K_s\frac{n^2}{\overline{r}^2}\sin^2f(\overline{r})\left(\frac{df(\overline{r})}{d\overline{r}}\right)^2\right\}\overline{r}~d\overline{r}~d\theta
\end{eqnarray}
In particular,
\begin{eqnarray}\label{35}
\left.\frac{\partial\mu_q}{\partial q}\right|_{q=1}
&=& 4qK_s\int\int\frac{n^2}{\overline{r}^2}\sin^2f(\overline{r})\left(\frac{df(\overline{r})}{d\overline{r}}\right)^2\overline{r}d\overline{r}d\theta\nonumber\\
&>& 0
\end{eqnarray}
But, if $f$ is a localized solution of eq.(\ref{28}), meaning that it remains suitably bounded as $r\rightarrow 0$ and as $r\rightarrow\infty$, it should be a stationary point of $\mu$, meaning that $\partial\mu_q/\partial q|_{q=1}=0$. 

It follows therefore that no localized solution of (\ref{28}) exists. A more rigorous statement of this property follows on from Derrick's theorem \cite{derrick}, which states that a necessary condition for vortex stability is that 
\begin{eqnarray}\label{36}
\left.\frac{\partial \mu}{\partial q}\right|_{q=1}&=&0
\end{eqnarray}
It is evident that (\ref{35}) does not satisfy this criterion.

In an attempt to fix this problem, we could add a ''mass'' term i.e. $K_v(1-\hat{\textbf{z}}.\phi)$, to the Lagrangian density, $\mathcal{L}_2$, where $\hat{\textbf{z}}$ is the direction of $\phi$ at $r=\infty$ (where $f(r)=0$). The Lagrangian density then becomes
\begin{eqnarray}\label{37}
\mathcal{L}_3=\mathcal{L}_2+K_v(1-\underline{n}.\hat{\underline{\phi}})
\end{eqnarray}
[This Lagrangian density corresponds to the baby Skyrmion model in equation (2.2), p.207 of \cite{piette}]. 

The kinetic term (in the case of a free particle) together with the Skyrme term in $\mathcal{L}_2$ are not sufficient to stabilize a baby Skyrmion, as the kinetic term in $2+1$ dimensions is conformally (scale) invariant and the baby Skyrmion can always reduce its energy by inflating indefinitely. This is in contrast to the usual Skyrme model, in which the Skyrme term prohibits the collapse of the $3+1$ soliton \cite{gisiger}. The mass term is added to limit the size of the baby Skyrmion. 

\section{Skyrmion Vortex with a Twist}
Instead of adding a mass term to stabilize the vortex, we will retain the baby Skyrme model Lagrangian (\ref{24}) but include a twist in the field, $g$, in (\ref{11}). That is, instead of choosing \cite{simanek}, \cite{cho}
\begin{equation}\label{38}
g=n\theta-\chi
\end{equation}
we choose
\begin{equation}\label{39}
g=n\theta+mkz
\end{equation}
where $mkz$ is the twist term, $m$ and $n$ are integers, $2\pi/k$ is the period in the $z$-direction.

The Lagrangian density (\ref{24}) then becomes 
\begin{eqnarray}\label{40}
\mathcal{L}_2
&=& \frac{1}{2\lambda^2}\left[\left(\frac{df}{dr}\right)^2+\sin^2f\left(\frac{n^2}{r^2}+m^2k^2\right)\right] \nonumber\\
&&+~2K_s\sin^2f\left(\frac{df}{dr}\right)^2\left(\frac{n^2}{r^2}+m^2k^2\right)
\end{eqnarray}
The value of the twist lies in the fact that in the far field, where $r\to\infty$ then $f\to0$, the Euler-Lagrange equations for $f$ for both $\mathcal{L}_3$ (without a twist) and $\mathcal{L}_2$ (with a twist) are formally identical to leading order, with $m^2k^2/\lambda^2$ in the twisted case playing the role of the mass coupling constant, $K_v$. So, it is expected that the twist term will act to stabilize the vortex just as the mass term does in $\mathcal{L}_3$.

On a physical level, the twist can be identified with a circular stress in the plane, perpendicular to the vortex string (which can be imagined e.g. as a rod aligned with the $z$-axis). The direction of the twist can be clockwise or counter-clockwise. In view of the energy-mass relation, the energy embodied in the stress term contributes to the gravitational field of the string, with the net result that the trajectories of freely-moving test particles differ according to whether they are directed clockwise or counter-clockwise around the string.

The Euler-Lagrange equation corresponding to the twisted baby Skyrmion vortex Lagrangian density (\ref{40}) reads
\begin{eqnarray}\label{41}
0
&=&\frac{1}{\lambda^2}\left[\frac{d^2f}{dr^2}+\frac{1}{r}~\frac{df}{dr}-\left(\frac{n^2}{r^2}+m^2k^2\right)~\sin f~\cos f\right] \nonumber\\
&&+~4K_s~\frac{n^2}{r^2}~\sin^2f\left(\frac{d^2f}{dr^2} -\frac{1}{r}\frac{df}{dr}\right) \nonumber\\
&&+~4K_s~m^2k^2\sin^2f\left(\frac{d^2f}{dr^2}+\frac{1}{r}\frac{df}{dr}\right)\nonumber\\
&&+~4\left(\frac{n^2}{r^2}+m^2k^2\right)K_s~\sin f~\cos f\left(\frac{df}{dr}\right)^2
\end{eqnarray}
It should be noted that the second Euler-Lagrange equation (\ref{27}) is satisfied identically if $g$ has the functional form (\ref{39}).

\section{Vortex Solution of the Twisted Baby Skyrme Equation}
Referring to eq.(\ref{41}), let us rewrite the Euler-Lagrange equation from $\mathcal{L}_2$, eq.(\ref{40}), i.e. the twisted baby Skyrme vortex equation as
\begin{eqnarray}\label{42}
\frac{d^2f}{dr^2}
&=&-\frac{(\varepsilon+\zeta~r^2)\sin f~\cos f}{r^2+(\varepsilon+\zeta r^2)\sin^2f}\left(\frac{df}{dr}\right)^2 \nonumber\\
&&-~\frac{1}{r}\left[\frac{r^2+(-\varepsilon+\zeta r^2)\sin^2f}{r^2+(\varepsilon+\zeta r^2)\sin^2f}\right]\frac{df}{dr} \nonumber\\
&&+~\frac{n^2(1+\frac{\zeta}{\varepsilon}r^2)\sin f\cos f}{r^2+(\varepsilon+\zeta r^2)\sin^2f}
\end{eqnarray}
$\varepsilon$ and $\zeta$ should be defined as $\varepsilon=2\lambda^2K_sn^2$ and $\zeta=4\lambda^2K_s(mk)^2$. Here, once again, $K_s$ is a positive coupling constant. 

Eq.(\ref{42}) will be solved numerically for different values of $\varepsilon$ and $\zeta$, starting with $f(0)=\pi$, $f'(0)=-a$ for different values of $a$. We need to solve (\ref{42}) numerically to calculate the minimum energy of the vortex. 

Let us use Runge-Kutta fourth order method for solving (\ref{42}). First, assume that
\begin{eqnarray}\label{43}
\frac{df}{dr}&=& v;~~~~~\frac{d^2f}{dr^2}=\frac{dv}{dr}
\end{eqnarray}
then
\begin{eqnarray}\label{44}
\frac{dv}{dr}
&=&g(r,~f,~v)\nonumber\\
&=&-~\frac{(\varepsilon+\zeta r^2)~\sin f~\cos f}{r^2+(\varepsilon+\zeta r^2)\sin^2f}v^2\nonumber\\
&&-~\frac{1}{r}\left[\frac{r^2+(-\varepsilon+\zeta r^2)\sin^2f}{r^2+(\varepsilon+\zeta r^2)\sin^2f}\right]v \nonumber\\
&&+~\frac{n^2(1+\frac{\zeta}{\varepsilon}r^2)\sin f\cos f}{r^2+(\varepsilon+\zeta r^2)\sin^2f}
\end{eqnarray}

Let us write down (\ref{43}), (\ref{44}) in iterative form as
\begin{eqnarray}\label{45}
v_{i+1}=v_i+\frac{dr}{6}(k_1+2k_2+2k_3+k4)
\end{eqnarray}
and
\begin{eqnarray}\label{46}
f_{i+1}=f_i+\frac{dr}{6}(l_1+2l_2+2l_3+l_4)
\end{eqnarray}
with
\begin{eqnarray}\label{47}
l_1=v_i;~k_1=g(r_i,~f_i,~v_i)
\end{eqnarray}
\begin{eqnarray}\label{48}
l_2
&=& v_i+\frac{dr}{2}k_1\nonumber\\
k_2
&=& g(r_i+\frac{dr}{2},f_i+\frac{dr}{2}l_1,v_i+\frac{dr}{2}k_1)
\end{eqnarray}
\begin{eqnarray}\label{49}
l_3
&=& v_i+\frac{dr}{2}k_2\nonumber\\
k_3
&=& g(r_i+\frac{dr}{2},f_i+\frac{dr}{2}l_2,v_i+\frac{dr}{2}k_2)
\end{eqnarray}
\begin{eqnarray}\label{50}
l_4
&=& v_i+dr~k_3\nonumber\\
k_4
&=& g(r_i+dr,f_i+dr~l_3,v_i+dr~k_3)
\end{eqnarray}
Eq.(\ref{45}) and (\ref{46}) give the numerical approximations to $df/dr$ and $f$ at the radius $r_{i+1}=(i+1)dr$. 

If a suitable vortex solution $f(r)$ of this equation exists, it should have a series expansion for $r<<1$ of the form
\begin{eqnarray}\label{51}
f=\pi + ar +br^3+...~~\text{if}~~n=1
\end{eqnarray}
or
\begin{eqnarray}\label{52}
f=\pi + ar^n +br^{n+2}+...~~\text{if}~~n\geq2
\end{eqnarray}
where $a<0$ and $b$ are constants, with
\begin{eqnarray}\label{53}
b
&=& \frac{a}{24\varepsilon}\frac{a^2(a^2\varepsilon-2)\varepsilon+3\zeta(1-2a^2\varepsilon)}{1+a^2\varepsilon}~~~\text{if}~~n=1
\end{eqnarray}
and
\begin{eqnarray}\label{54}
b
&=& \frac{a}{4\varepsilon}\frac{n^2\zeta}{n+1}~~~\text{if}~~n\geq 2.
\end{eqnarray}

To determine the asymptotic form of the solution for $r>>1$, note first that if $f<<1$ the field equation (\ref{42}) becomes, to linear order in $f$,
\begin{eqnarray}\label{55}
\frac{d^2f}{dr^2}
&=& -\frac{1}{r}\frac{df}{dr}+n^2(\zeta/\varepsilon)f
\end{eqnarray}
$u$ should be defined as $u=\sqrt{(\zeta/\varepsilon) r}$, this equation reads
\begin{eqnarray}\label{56}
u^2\frac{d^2f}{du^2}+u\frac{df}{du}-u^2f
&=& 0
\end{eqnarray}
which is just the modified Bessel equation of order 0. 

Any solution to this equation which vanishes as $u\rightarrow\infty$ is proportional to the modified Bessel function $K_0(u)$, which is known to have asymptotic form
\begin{eqnarray}\label{57}
K_0(u)
&\approx& \sqrt{\frac{\pi}{2u}}e^{-u}+O(u^{-3/2}e^{-u})
\end{eqnarray}
for $u>>1$. So, we expect that any solution to (\ref{42}) which vanishes as $r\rightarrow\infty$ will have the asymptotic form 
\begin{eqnarray}\label{58}
f(r)
&=& e^{-n\sqrt{(\zeta/\varepsilon) r}}(Ar^{-1/2} +Br^{-3/2}+...)
\end{eqnarray}
where $A$ is a positive constant. 

Substituting this asymptotic expansion into (\ref{42}) shows that
\begin{eqnarray}\label{59}
B
&=& A\frac{4n^2-1}{8n\sqrt{\zeta/\varepsilon}}
\end{eqnarray}
So, the asymptotic form of $f$ is fixed by a single constant $A$, which must be chosen so that $f$ and $f'$ in the asymptotic ($r>>1$) solution match $f$ and $f'$ in the small-distance ($r<<1$) solution at some intermediate distance. (Both solutions need to be obtained by numerical integration.) 

In contrast to the previous (twist-free) case, we do expect solutions to (\ref{42}) to exist and to be energetically stable, as the argument below demonstrates. Following the same procedure as in previous chapter, suppose that $f(r)$ is a solution of equation (\ref{42}), let $q$ be any positive constant, and define $f_q(r)=f(qr)$. Substituting $f_q$ in place of $f$ in the formula $\mu=2\pi\int(-\mathcal{L}_2)r~dr$ gives a value of $\mu$ which again depends on $q$:
\begin{eqnarray}\label{60}
\mu_q
&=& \frac{\pi}{\lambda^2}\int_0^\infty\left[\left(\frac{df_q}{dr}\right)^2+n^2\left(\frac{1}{r^2} +\frac{\zeta}{\varepsilon}\right)\sin^2f_q\right.\nonumber\\
&&\left.+~\left(\frac{df_q}{dr}\right)^2\left(\zeta+\frac{\varepsilon}{r^2}\right)\sin^2f_q \right]r~dr
\end{eqnarray}

Noting that
\begin{eqnarray}\label{61}
\frac{df_q}{dr}
&=& qf'(qr)
\end{eqnarray}
as before, and replacing $r$ with $\bar{r}=qr$ gives
\begin{eqnarray}\label{62}
\mu_q
&=& \frac{\pi}{\lambda^2}\int_0^\infty\left[\left(\frac{df(\bar{r})}{d\bar{r}}\right)^2+n^2\left(\frac{1}{\bar{r}^2} +\frac{\zeta}{\varepsilon q^2}\right)\sin^2f(\bar{r})\right.\nonumber\\
&&+~\left.\left(\frac{df(\bar{r})}{d\bar{r}}\right)^2\left(\zeta+\frac{\varepsilon q^2}{\bar{r}^2}\right)\sin^2f(\bar{r}) \right]~\bar{r}~d\bar{r}
\end{eqnarray}
So,
\begin{eqnarray}\label{63}
\frac{\partial\mu_q}{\partial q}|_{q=1}
&=& \frac{2\pi}{\lambda^2}\left[-\int_0^\infty \frac{n^2\zeta}{\varepsilon}\sin^2f(\bar{r})\bar{r}~d\bar{r}  \right.\nonumber\\
&&\left.+\int_0^\infty\left(\frac{df(\bar{r})}{d\bar{r}}\right)^2\frac{\varepsilon}{\bar{r}}\sin^2f(\bar{r})~d\bar{r}\right]
\end{eqnarray}
If a localised solution of eq.(\ref{42}) exists, it must have $\partial\mu_q/\partial q|_{q=1}=0$, so any solution $f(r)$ will satisfy
\begin{eqnarray}\label{64}
\varepsilon\int_0^\infty\left(\frac{df}{dr}\right)^2\frac{1}{r}\sin^2f(r)dr
=\frac{n^2\zeta}{\varepsilon}\int_0^\infty\sin^2f(r)r~dr\nonumber\\
\end{eqnarray}

A necessary condition for a stationary (time-independent) solution of the kind we are considering here to be stable is that $\partial^2\mu_q/\partial q^2|_{q=1}\geq0$ (since otherwise it would be energetically favourable for the to either dilate or shrink uniformly while radiating energy: a result often referred to as Derrick's Theorem \cite{derrick}). Here
\begin{eqnarray}\label{65}
\left.\frac{\partial^2\mu_q}{\partial q^2}\right|_{q=1}
&=& \frac{2\pi}{\lambda^2}\left[3\int_0^\infty\frac{n^2\zeta}{\varepsilon}\sin^2f(\bar{r})~\bar{r}~d\bar{r} \right.\nonumber\\
&&\left.+~\int_0^\infty\left(\frac{df(\bar{r})}{d\bar{r}}\right)^2\frac{\varepsilon}{\bar{r}}\sin^2f(\bar{r})~d\bar{r}\right] \nonumber\\
&=& \frac{8\pi}{\lambda^2}\frac{n^2\zeta}{\varepsilon}\int_0^\infty \sin^2f(\bar{r})~\bar{r}~d\bar{r}
\end{eqnarray}
which is clearly non-negative provided that $\zeta/\varepsilon \geq 0$. So, all solutions of the twisted baby Skyrme vortex equation - if any exist - are guaranteed to be stable.

We have attempted to numerically integrate equation (\ref{42}) for the range of parameters namely $n=1$, $\varepsilon=0.05$ and $\zeta$ from 0 to $4\times 10^{-8}$. The best way to do this, we have found, is to write (\ref{42}) in the form
\begin{eqnarray}\label{66}
f''
&=& \zeta r\frac{-f'^2r(\cos f/\sin f) -f'+r(\cos f/\sin f)/\varepsilon}{(r^2/\sin^2f)+(\varepsilon +\zeta r^2)} \nonumber\\
&&+~\varepsilon r^{-1}f'\frac{1-rf'\cos f/\sin f}{(r^2/\sin^2f)+(\varepsilon+\zeta r^2)} \nonumber\\
&&+\frac{1}{\sin f}\frac{\cos f-(rf'/\sin f)}{(r^2/\sin^2f)+(\varepsilon+\zeta r^2)}
\end{eqnarray}
and then integrate it from $r\approx 10^{-2}$ with
\begin{eqnarray}\label{67}
f
&=& \pi +ar +\frac{a}{24\varepsilon}\frac{a^2(a^2\varepsilon -2)\varepsilon +3\zeta(1-2a^2\varepsilon)}{1+a^2\varepsilon}r^3
\end{eqnarray}
out to $r=(2\ln 2)/\sqrt{\zeta/\varepsilon}\approx 3000$ using 4th-order Runge-Kutta with 35000 steps. 

To continue out to $r\rightarrow\infty$, we define $x=e^{-\sqrt{(\zeta/\varepsilon) r}/2}$ and rewrote (\ref{42}) in the form
\begin{eqnarray}\label{68}
f_{xx}
&=& f_x^2\frac{-[(\ln x)^{-2}+4]\zeta\sin f\cos f}{4+[(\ln x)^{-2}+4]\zeta\sin^2f} \nonumber\\
&&+~\frac{x^{-1}f_x}{|\ln x|}\frac{4+[-(\ln x)^{-2}+4]\zeta\sin^2f}{4+[(\ln x)^{-2}+4]\zeta\sin^2f} -x^{-1}f_x \nonumber\\
&&+4x^{-2}\frac{[(\ln x)^{-2}+4]\sin f\cos f}{4+[(\ln x)^{-2}+4]\zeta\sin^2f}
\end{eqnarray}
Then (from the first point above) the solution should have the asymptotic form
\begin{eqnarray}\label{69}
f
&=& Cx^2(|\ln x|^{-1/2} +\frac{3}{16}|\ln x|^{-3/2})
\end{eqnarray}
and
\begin{eqnarray}\label{70}
f_x
&=& Cx(2|\ln x|^{-1/2} +\frac{7}{8}|\ln x|^{-3/2})
\end{eqnarray}
for some constant $C>0$. 

Starting from $x=2.5\times 10^{-4}$, we integrated the equation out to $x=1/2$ (using 4th-order Runge-Kutta with 5000 steps) and attempted to match the two values of $f$, plus their derivatives $f_r=-\frac{1}{2}\sqrt{\zeta/\varepsilon x}f_x$ by varying the constants $a$ and $C$. 

We also calculated the mass per unit length for the matched solutions from the formula
\begin{eqnarray}\label{71}
\mu 
&=& 2\pi\int_0^{(2\ln 2)/\sqrt{\zeta/\varepsilon}}\left[\left(\frac{df}{dr}\right)^2 +\sin^2f(r^{-2}+\zeta/\varepsilon) \right.\nonumber\\
&&\left.+~\left(\frac{df}{dr}\right)^2\sin^2f(\varepsilon r^{-2}+\zeta)\right]r~dr \nonumber\\
&&+~2\pi\int_0^{1/2}\left\{f_x^2 +(\sin f/x)^2[(\ln x)^{-2} +4] \right.\nonumber\\
&&\left.+~\frac{1}{4}\zeta f_x^2\sin^2f[(\ln x)^{-2}+4]\right\}|\ln x|x~dx \nonumber\\
\end{eqnarray}
and the two integrals
\begin{eqnarray}\label{72}
I_1
&=& 2\pi\varepsilon\int_0^{(2\ln 2)/\sqrt{\zeta/\varepsilon}}\left(\frac{df}{dr} \right)^2\frac{1}{r}\sin^2f(r)dr \nonumber\\
&&+~\frac{\pi}{2}\zeta\int_0^{1/2}f_x^2\sin^2f|\ln x|^{-1} x~dx
\end{eqnarray}
\begin{eqnarray}\label{73}
I_2
&=& \frac{2\pi\zeta}{\varepsilon}\int_0^{(2\ln 2)/\sqrt{\zeta/\varepsilon}}\sin^2f(r)r~dr \nonumber\\
&&+~8\pi\int_0^{1/2}(\sin f/x)^2|\ln x|^{-1}x~dx
\end{eqnarray}
which should equal whenever a solution exists. 

\section{Results}
 The results we have generated so far (after much trial and error) for $\varepsilon = 0.05$ are:
\begin{center}
\begin{tabular}{ |c|c|c|c|c| } 
 \hline
 $\zeta$ & $a$ & $C$ & $\mu$ \\ 
 \hline
  10$^{-8}$ & -0.141482 & 0.0111948 & 25.14166  \\ 
 \hline
 2$\times$10$^{-8}$ & -0.166341 & 0.0134676 & 25.14509  \\
 \hline 
 3$\times$10$^{-8}$ & -0.182811 & 0.0150025 & 25.14767  \\
 \hline
 4$\times$10$^{-8}$ & -0.195450 & 0.0162065 & 25.14983  \\
 \hline
\end{tabular}
\end{center}

\begin{center}
\begin{tabular}{ |c|c|c|c|c|c| } 
 \hline
 $\zeta$ & $I_1$ & $I_2$ \\ 
 \hline
  10$^{-8}$ & 0.004194756 & 0.004194685 \\ 
 \hline
 2$\times$10$^{-8}$ & 0.005799584 & 0.005799716 \\
 \hline 
 3$\times$10$^{-8}$ & 0.007006055 & 0.007006095 \\
 \hline
 4$\times$10$^{-8}$ & 0.008009381 & 0.008009438 \\
 \hline
\end{tabular}
\end{center}

\section{Discussions and Conclusions}
The results are just as expected. If $\varepsilon$ and $\zeta$ are fixed, there is only one value of $a$ (and one value of $C$) for which a solution exists. If $|a|$ is too large, the value of $f$ eventually become negative. If $|a|$ is too small, $f$ reaches a finite positive minimum value at a finite value of $r$, and then starts increasing again. As can be seen, all the trends are simple, and $\mu$ increases roughly linearly with $\zeta$.

\section{Acknowledgment}
MH thank to UBD GRS Scholarships for supporting this research. 
\\

\end{document}